\documentclass{article}
\usepackage{graphicx, float,amsmath,subfigure,authblk,fancyhdr,lastpage}
\pdfoutput=1 
\usepackage{geometry}
\title{\bf A nonlinear evolution equation for pulsating detonations using Fickett's model with chain branching kinetics}

\author{A. Bellerive \& M.I. Radulescu}
\affil{\emph{\small Department of Mechanical Engineering, University of Ottawa, Ottawa, Ontario, Canada K1N6N5}}

\date{}

\newcommand*\diff{\mathop{}\!\mathrm{d}}

\begin{document}
    \maketitle
\begin{abstract}
The detonation wave stability is addressed using Fickett's equation, i.e., the reactive form of Burgers' equation. This serves as a simple analogue to the reactive Euler equations, permitting one to gain insight into the nonlinear dynamics of detonation waves.  Chemical kinetics were modeled using a two-step reaction with distinct induction and reaction zones.  An evolution equation for the detonation structure was derived using the method of matched asymptotics for large activation energy and slow rate of energy release.  While the first order solution was found unconditionally unstable, the second order evolution equation predicted both stable and unstable solutions.  The neutral stability boundary was found analytically, given by $\chi=4$, where $\chi$ is the product of activation energy and the ratio of induction to reaction time.  This reproduces accurately what has been previously established for the reactive Euler equations and verified experimentally.  The evolution equation also captures stable limit cycle oscillations in the unstable regime and offers unique insight into the instability mechanism. The mechanism amplifying the perturbations lies within the induction zone, where the Arrhenius-type rate equation provides a large change in induction times for small perturbations. The mechanism attenuating the perturbations arises from acoustic effects, which delays the amplification of the shock front. The longer the detonation wave, the more time it takes for the amplification from the reaction zone to reach the shock front, creating gradients that counter-act the amplification from the flame acceleration. The results agree with direct numerical simulation, as well as recovering many similarities with the reactive Euler equation.
\end{abstract}
\thispagestyle{fancy}
\section{Introduction}
Detonations are self-sustained supersonic waves consisting of a shock front that is driven by a volumetric expansion induced by chemical reactions of the shocked material. 
Zeldovich, Von Neuman and D\"{o}ring discovered the idealized steady structure and is known as the ZND wave. The minimum sustainable steady detonation speed is known as the Chapman-Jouguet (CJ) detonation velocity, where the detonation velocity is simply the speed at which the equilibrium or fully reacted zone flow is exactly sonic relative to the lead shock. In practice, detonations are rarely stable and have two main instability modes: one-dimensional pulsating or multi-dimensional cellular. The natural tendency of a reactive mixture to detonate has been correlated with stability characteristics\cite{radulescu_2002}. It has also been shown that unstable detonations have wider limits\cite{radulescu_2002,moen_1986} than stable detonations. Determining the parameters that control detonation stability is of particular interest. The product of the ratios of induction to reaction time and the activation energy,
\begin{equation}
\chi=\frac{t_{ind}}{t_{reac}}\frac{E_a}{RT_s},
\end{equation} 
has been identified in the stability analyses of self-propagating detonation waves\cite{short_2003,radulescu_2005,leung_2010}. The physical mechanism responsible for changes of the shock-ignition regime when $\chi$ is varied remain unclear.

There are typically three different approaches for analyzing stability problems: linear stability analysis, direct numerical simulation, and  asymptotic modeling. There is a wealth of literature for each approach, a recent review by Ng and Zhang offers the current state of knowledge\cite{zhang_2011}. Linear stability permits one to determine the parameters controlling stability but lacks in identifying the mechanisms. Direct numerical simulation can yield accurate simulations for nonlinear systems in 1-D and 2-D, but recovering the explanation why some detonation are stable and others are not is difficult. Finally, asymptotic modeling offers a better understanding of the mechanisms behind detonation stability, but are restricted by their limits of derivation. 

Such asymptotic methods have been used on the Reactive Euler equations to achieve analytical solutions by many researchers. The typical strategy is formulated using a combination of the following: the limit of large activation energy\cite{buckmaster_1988,short_2001}, the Newtonian limit\cite{short_1996}, overdriven detonation limit\cite{clavin_1996}, and/or the limit of weak heat release\cite{he_2000,short_1999,short_2002}. Nevertheless, such methods become very complex and limit high order analysis of the reactive Euler equations due to strong nonlinearities.

Detonation analogues have been introduced by Fickett\cite{fickett_1979,fickett_1985}, and by Majda and Rosales\cite{majda_rosales_1983} to better understand detonations. They are both a variation of Burgers' equation\cite{burgers_1948} with an added energy release term. The former has two families of waves, forward traveling pressure waves and energy release along the particle path. The model proved useful in direct numerical simulation and asymptotic modeling of piston initiated detonations\cite{tang_2013b}, and direct numerical simulation of self-sustained detonations leading to chaotic oscillation\cite{radulescu_2011}. The Majda-Rosales model has forward traveling waves and infinitely fast backward traveling waves, i.e., instantaneous. Direct numerical simulation has also shown chaotic oscillations\cite{kasimov_2013,faria_2014}. 

The present work focuses on asymptotic modeling for pulsating CJ detonations in Fickett's equation\cite{fickett_1979,fickett_1985}. The simplicity of Fickett's model can serve to study the stability mechanism without the complexity of the strong nonlinearities of the reactive Euler equations while still reproducing the complex detonation dynamics\cite{radulescu_2011}. 

The reaction model is a two-step, induction-reaction, studied in detail by Short \& Sharpe\cite{short_2003}, Ng et al.\cite{radulescu_2005}, and Leung, Radulescu \& Sharpe\cite{leung_2010} for the reactive Euler equations, and by Radulescu and Tang\cite{radulescu_2011,tang_2013b} for the Fickett model. Analytical solutions to the pulsating detonations and stability boundary are sought using asymptotic expansions at the limit of high activation energy.

The paper is organized as follows. First, the model is presented and an oscillator equation is derived for the lead shock perturbation speed, which takes the form of a second order nonlinear ODE. The dynamics predicted by the oscillator are then discussed. The predicted neutral boundary stability is obtained in closed form in terms of $\chi$ and compared with numerical solutions. The mechanism of instability is then discussed from the analysis of the oscillator.  

\section{The Model}
The detonation wave is modeled with Fickett's equations, an analogue to the compressible reactive Euler equations. The model is of the same form as the inviscid Burgers' equation\cite{burgers_1948}, $u_t+u u_x=0$, with an added energy release term in the reaction layer. The following equations describing the model are the conservation, state, and reaction rate:

\begin{eqnarray}
\label{conservation.eq}		
\frac{\partial\rho}{\partial t}+\frac{\partial  p}{\partial x}=0, 	
\\\label{pressure.eq}		 
p=\frac{1}{2}(\rho^2+\lambda_r Q), 				  
\\\label{rate.eq}
\frac{\partial \lambda_{(i,r)}}{\partial t}=r(\rho,\lambda_{(i,r)}),			
\end{eqnarray}
where $\rho$ is analogous to density, $p$ is the pressure, $Q$ is the heat release parameter, $r$ the reaction rate, and $\lambda_i$ and $\lambda_r$ will denote the induction and reaction zone progress variables, respectively.  

A large majority of chemical reactions begin with chain initiation followed by a sequence of chain-branching, chain-recombination and chain termination steps. The modeled detonation structure is assumed to have a two-step reaction: chain-initiation followed by chain-branching/recombination and chain termination. The former is assumed to be thermally neutral, i.e., no heat release, and the latter to be exothermic and independent of the thermodynamic state, i.e., independent of $\rho$ for this model. The chain-initiation dynamics in the induction zone are controlled by a density-sensitive Arrhenius form of the reaction rate,
\begin{equation}
\label{rate.induction.eq} 
\frac{\partial \lambda_i}{\partial t}=r_i=-K_i \exp(\alpha(\frac{ \rho}{2  \rho_{cj}}-1)),
\end{equation}
where $\rho_{cj}$, is the Chapman-Jouguet density, $K_{i}$ is the constant of reaction and $\alpha$ is the activation energy. $K_i$ controls the induction zone length and $\alpha$ controls the sensitivity to changes in density. The progress variable, $\lambda_i$, in Eq.~\eqref{rate.eq}, is equal to $1$ at the shock front and $0$ at the end of the induction zone. The heat release is then triggered when $\lambda_i$ is equal to 0. The energy release rate is taken to be of the form
\begin{equation}
\label{rate.reaction.eq}
\frac{\partial \lambda_r}{\partial t}=r_r=K_r(1-\lambda_r)^\nu,
\end{equation}
where $K_r$ is the reaction constant, and $\nu$ the reaction order. The reaction order is assumed to be such that $1/2 \le \nu < 1$. A reaction order of non-unity is used to approximate several steps of chain-branching and chain termination that is typically found. $\lambda_r=1$ signals the rear equilibrium or sonic point of the detonation wave.
\section{Steady Structure}
\begin{figure} 
\centering
\includegraphics[width=0.6\linewidth]{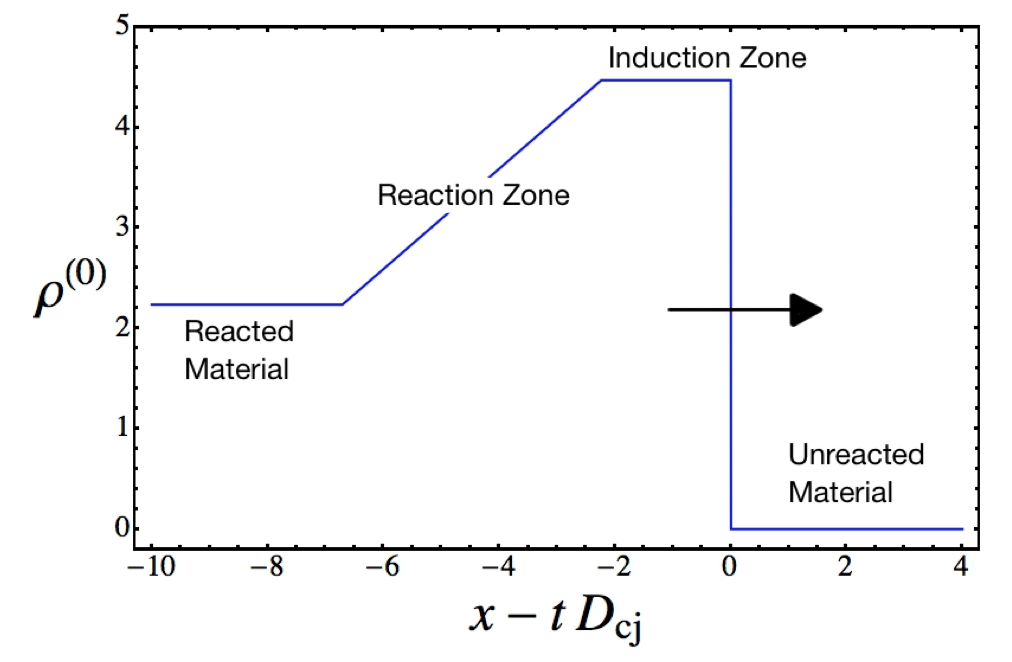}
   \caption{Steady detonation structure in the shock attached frame of reference traveling at the Chapman-Jouguet detonation velocity $D_{cj}$, for $K_i=1, K_r=2$ and the heat release parameter $Q=5.$}
   \label{structure.fig}
\end{figure}

To obtain the steady structure of the traveling wave solution, the spatial variable is changed to a shock attached frame with $\zeta=x-Dt$, where $D$ is the steady detonation speed. The conservation~\eqref{conservation.eq}, state~\eqref{pressure.eq}, and rate~\eqref{rate.eq} equations in the steady shock attached frame are then given by
\begin{equation}
\label{Ficketts.expanded.shock.attached}
\frac{\diff}{\diff \zeta}\bigg(\frac{\rho^2}{2}-D\rho+\frac{1}{2}\lambda_rQ\bigg)=0,
\end{equation}
\begin{equation}
\frac{\diff\lambda}{\diff\zeta}=-\frac{r}{D}.
\end{equation}
Integrating these equations yields the analytical results for the steady wave structure in the induction zone ($\zeta_{ind}<\zeta<0$)
\begin{equation}
\label{stead.rho.ind}
\rho_{ind}=2 D_{cj},
\end{equation}
\begin{equation}
\lambda_i=\frac{K_i}{D_{cj}}\zeta+1,
\end{equation}
\begin{equation}
\label{steady.zeta.ind}
\zeta_{ind}=-\frac{D_{cj}}{K_i},
\end{equation}
and in the reaction zone ($\zeta_{cj}<\zeta<\zeta_{ind}$)
\begin{equation}
\label{steady.rho.reac}
\rho_{r}=D_{cj}(1+\sqrt{1-\lambda_r}),
\end{equation}
\begin{equation}
\lambda_r=1-[1+(1-\nu)\frac{K_r}{D_{cj}}(\zeta-\zeta_{ind})]^{\frac{1}{1-\nu}},
\end{equation}
\begin{equation}
\zeta_{cj}=\zeta_{ind}-\frac{D_{cj}}{K_r(1-\nu)}.
\end{equation}
with $\zeta_{ind}$ the end of the induction zone, and $\zeta_{cj}$ at the end of the reaction zone at CJ velocity. The CJ velocity is found with the same solution as Burgers' equation\cite{fickett_1979},
\begin{equation}
D=\frac{p_{\text{reac}}-p_o}{\rho_{\text{reac}}-\rho_\text{o}},
\end{equation}
\begin{equation}
\label{steady.Dcj}
D_{cj}=\sqrt{Q},
\end{equation}
where subscript "reac" is the fully reacted state, and subscript "o" is the undisturbed material. Illustrated in Figure~\ref{structure.fig} is an example steady detonation wave.
\section{Nonlinear Evolution Equation}
The evolution equation is derived assuming a high activation energy by setting $\alpha=\epsilon^{-1}$ with $\epsilon\ll1$, a longer reaction to induction time with $K_r/K_i\ll1$, and a slow pulsating evolution time with $\tau=\epsilon t$, with $t$ scaled to the induction delay time. The unsteady shock attached frame of reference spatial variable will be of the form 
\begin{equation}
n=x-D(t)t,
\end{equation} 
with the shock path
\begin{equation}
 D(t)=D_{cj}+h_{t}, 
\end{equation}
where $h$ is the position of the perturbed shock relative to the steady CJ detonation.

The variables are scaled to the steady post shock Chapman-Jouguet detonation values shown in the previous section. The non-dimensional variables are
\begin{equation}
\label{scales}
\begin{split}
&\tilde\rho=\frac{\rho}{\rho_s}, \quad \tilde n=\frac{ n}{ n_s}, \quad \tilde t=\frac{ t}{ n_s / D_s},\\ & \tilde K=\frac{ Kr}{ Ki}, \quad  K_i=1,
\end{split}
\end{equation}
where an overtilde ($\tilde{\;} $) denotes non-dimensional quantities and subscript $s$ denotes values in the immediate post shock steady detonation wave. Accordingly, the density is scaled to the induction zone's steady density~\eqref{stead.rho.ind}. The length $n$ is scaled with the induction zone's length~\eqref{steady.zeta.ind}. Time is scaled so 1 unit of $t$ is the time for a particle to pass through the induction zone, with $D_s$ the Chapman-Jouguet velocity~\eqref{steady.Dcj}. Fickett's equation in a shock traveling frame of reference scaled to immediate post-shock values in the steady wave may be written as
	\begin{equation}
	\label{governing.n.eq}\frac{\partial}{\partial n}\bigg(\rho^2-\big(1+\epsilon h_\tau\big)\rho+\frac{\lambda_r}{4}\bigg)=-\epsilon\frac{\partial \rho}{\partial \tau},
	\end{equation}
where the overtildes ($\tilde{\;} $) on non-dimensional variables are removed for simplicity, $\epsilon$ is the inverse activation energy, subscript $\tau$ is the slowed time derivative, $h_\tau$ is the shock front velocity position relative to the steady CJ detonation, and $1/4$ is the scaled heat release parameter. Similarly, the reaction rate equation \eqref{rate.eq} is given by
\begin{equation}
\label{raten} 
\frac{\partial \lambda_{(i,r)}}{\partial n}=\frac{1}{1+\epsilon h_\tau}\bigg[\epsilon \frac{\partial \lambda_{(i,r)}}{\partial \tau} - r_{(i,r)}\bigg].
\end{equation}
The scaled reaction rates for the induction and reaction zone, respectively, are
\begin{equation}
\label{rate.ind.normalized.eq}
r_i=-\exp\Big(\frac{1}{\epsilon}(\rho-1)\Big),
\end{equation}
\begin{equation}
\label{rate.r.normalized.eq}
r_r=K(1-\lambda_r)^\nu,
\end{equation}
where $r_i$ is scaled to the induction time, $K$ is the ratio of induction to reaction time, and $\nu$ is the reaction order ($\nu=1/2$ for this analysis). The variables are expanded about the steady solutions~\eqref{stead.rho.ind} and \eqref{steady.rho.reac}
\begin{equation}
\label{induction.density.eq}
\begin{split}
&\rho=\rho^{(0)}+\epsilon \rho^{(1)}+ \epsilon^2 \rho^{(2)}+O(\epsilon^3)+...,\\
&\lambda_r=\lambda_r^{(0)}+\epsilon \lambda_r^{(1)}+O(\epsilon^2)+...
\end{split}
\end{equation}
These expansions are substituted into \eqref{governing.n.eq} and \eqref{raten}, and integrated across the detonation wave for $O(1)$, $O(\epsilon)$, and $O(\epsilon^2)$. 

The $O(1)$ solution, is simply the steady Chapman-Jouguet detonation wave with a ZND profile shown in the previous section, see Fig.~\ref{structure.fig}. Next, $O(\epsilon)$ are integrated across the shockwave, followed by integrating to the end of the induction zone. The end of the induction zone is found by integrating~\eqref{rate.ind.normalized.eq} with~\eqref{raten}, and now defines a new function ${n=F(\tau)}$:
\begin{equation}
\label{induction.length}
\begin{split}
& F(\tau)=-\exp(-\rho^{(1)}),\\
& \rho^{(1)}=h_\tau.
\end{split}
\end{equation}
Due to the Arrhenius-type reaction rate~\eqref{rate.ind.normalized.eq}, an $O(\epsilon)$ change in $h_\tau$ results in an $O(1)$ change in the length of the induction zone~\eqref{induction.length}. Subsequently, the solution at the end of the induction zone is used as the boundary for the onset of reaction. The density is then integrated across the reaction zone to fully reacted material. A singular solution for $\rho^{(1)}$ at $\lambda_r=1$ (fully reacted) is found due to the sonic value of the fully reacted material in $O(1)$. Consequently, to achieve equilibrium at the fully reacted point the following condition is met
\begin{equation}
\label{evolution.O1.eq} 
3h_\tau-e^{-h_\tau}h_{\tau\tau}=0,
\end{equation}
the $O(\epsilon)$ evolution equation, which admits exponential solutions, i.e., accelerating or decelerating detonations waves for positive ($h_\tau|_{\tau=0}>0$) and negative ($h_\tau|_{\tau=0}<0$) initial perturbations, respectively. The same result was found in the reactive Euler equations for the same two-step reaction model\cite{short_2001}, and for a one-step reaction model\cite{buckmaster_1988}.

The dampening effects, attenuating or saturating the perturbations, are achieved at higher order, $O(\epsilon^2)$. By following the same integration for higher order, a similar singular solution occurs for $\rho^{(2)}$. Combining the remaining terms with~\eqref{evolution.O1.eq}, leads to the full nonlinear evolution equation: 
\begin{equation}
\begin{split}
\label{evolution.eq}
&\alpha_1 h_{\tau\tau\tau} e^{-h_\tau}+(\alpha_1+\alpha_2 e^{-h_\tau})h_{\tau\tau}+\alpha_3 h_\tau
+(\alpha_4+\alpha_5 e^{-h_\tau} ) e^{-h_\tau} h^2_{\tau\tau}+\\& \alpha_5 h^2_\tau
+\alpha_6 e^{-h_\tau} h_\tau h_{\tau\tau}=0,
\end{split}
\end{equation}
where the constants are given by
\begin{equation}
\label{constants}
\begin{aligned}
\alpha_1 =& \frac{\epsilon}{2K(1-\nu)}, \quad & \alpha_2 = \epsilon-\frac{1}{4}, \\ 
\alpha_3 =& \frac{3}{4}, \quad &\alpha_4= -\alpha_1, \\ 
\alpha_5 =& \frac{\epsilon}{4}, \quad &\alpha_6 = - \frac{3 \epsilon}{4}.
\end{aligned}
\end{equation}
\section{Numerical Solutions to the Evolution Equation}
The nonlinear evolution equation~\eqref{evolution.eq} is solved numerically. Due to the stiff nature of the nonlinear ODE, a backwards derivation formula with Newton iterations\cite{cvode_code_1996} is used. To simulate shock perturbations, initial conditions are taken as $h_{\tau}=0$ and $h_{\tau\tau}=0.1$. The evolution equation~\eqref{evolution.eq} predicts both stable and unstable solutions. The former decays to the steady CJ detonation wave. Due to nonlinearities, the latter's amplitude of oscillation amplifies exponentially until reaching a limit-cycles.
\begin{figure}[ht]
	\centering
	\subfigure[Shock front density history.]{
	\includegraphics[width=0.45\linewidth]{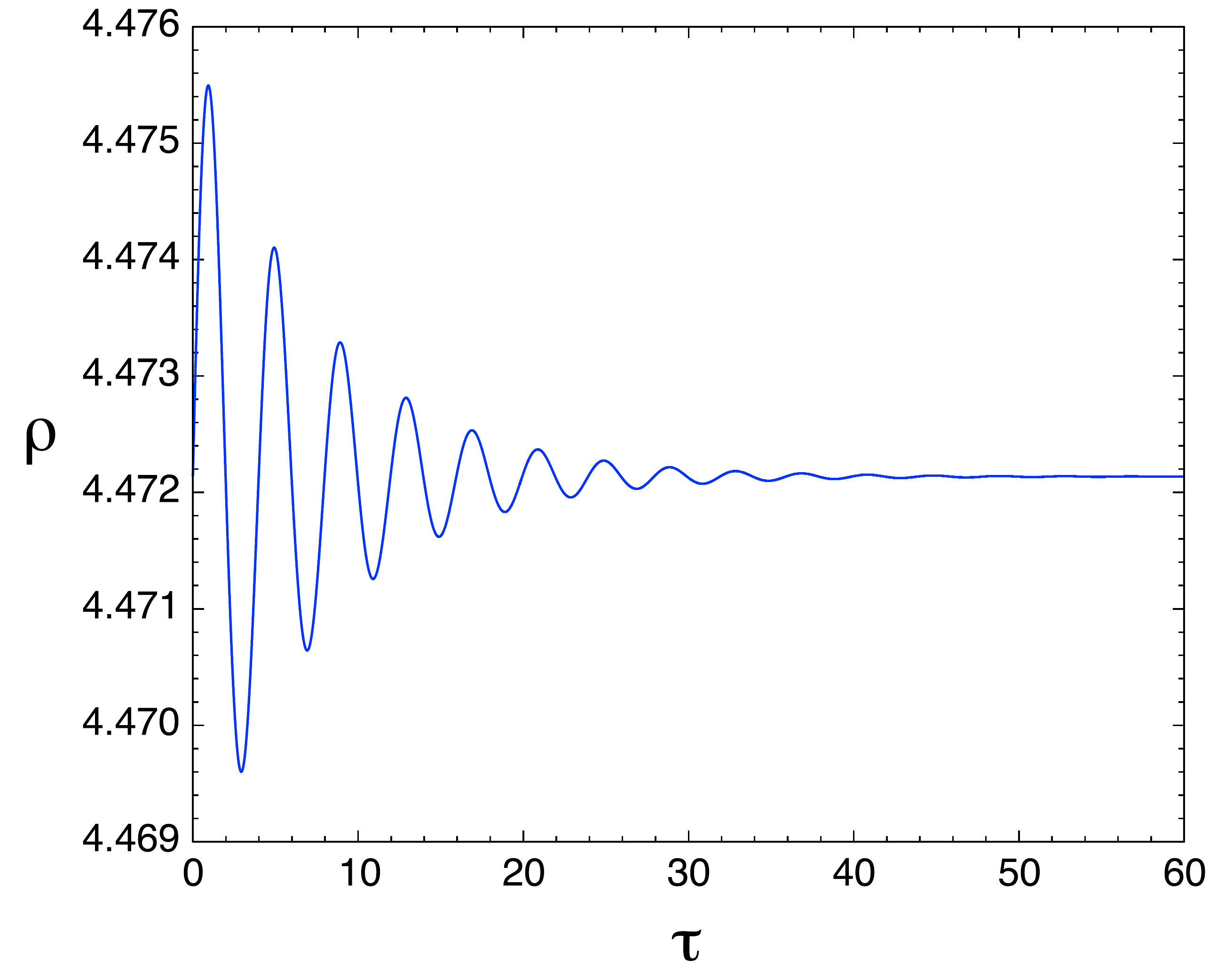}}
	\subfigure[Shock front acceleration vs. relative velocity.]{
	\includegraphics[width=0.45\linewidth]{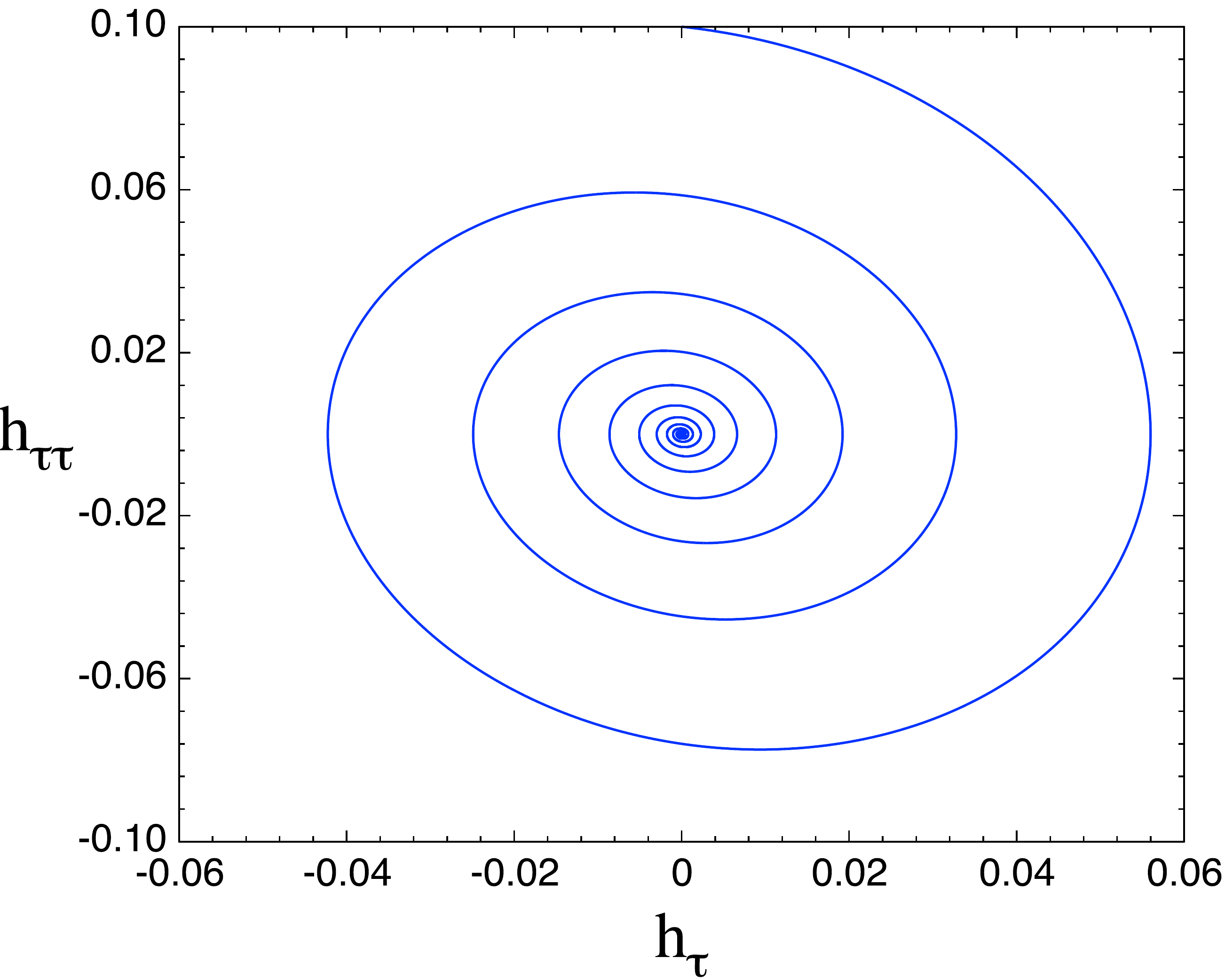}}
	\caption{(a) Stable solution with $K=0.1$ and $\alpha=33$ and (b) shock speed relative to steady wave phase portrait.}
	\label{fig:stable}
\end{figure}
\begin{figure}[ht]
	\centering
	\subfigure[Shock front density history.]{
	\includegraphics[width=0.45\linewidth]{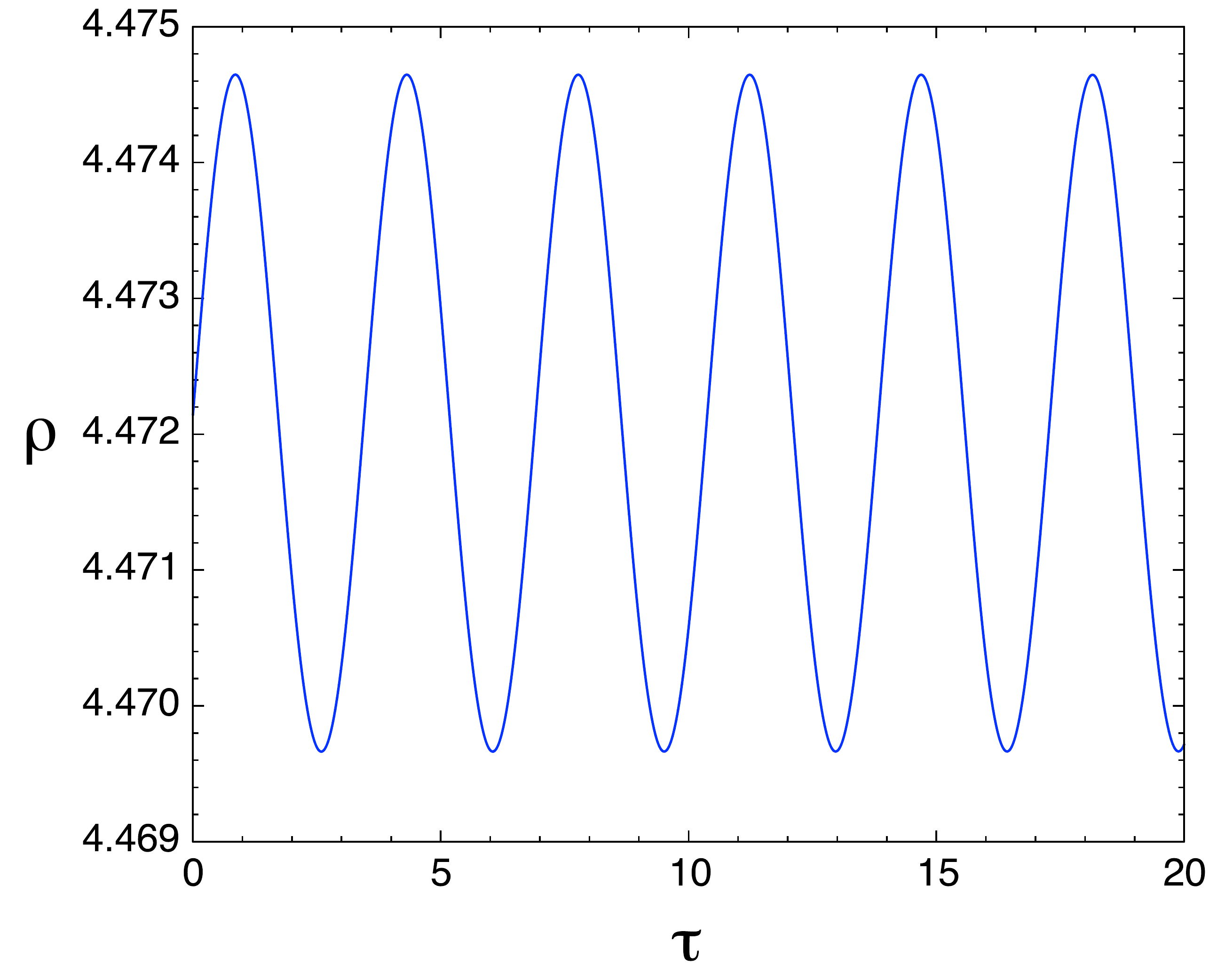}}
	\subfigure[Shock front acceleration vs. relative velocity.]{
	\includegraphics[width=0.45\linewidth]{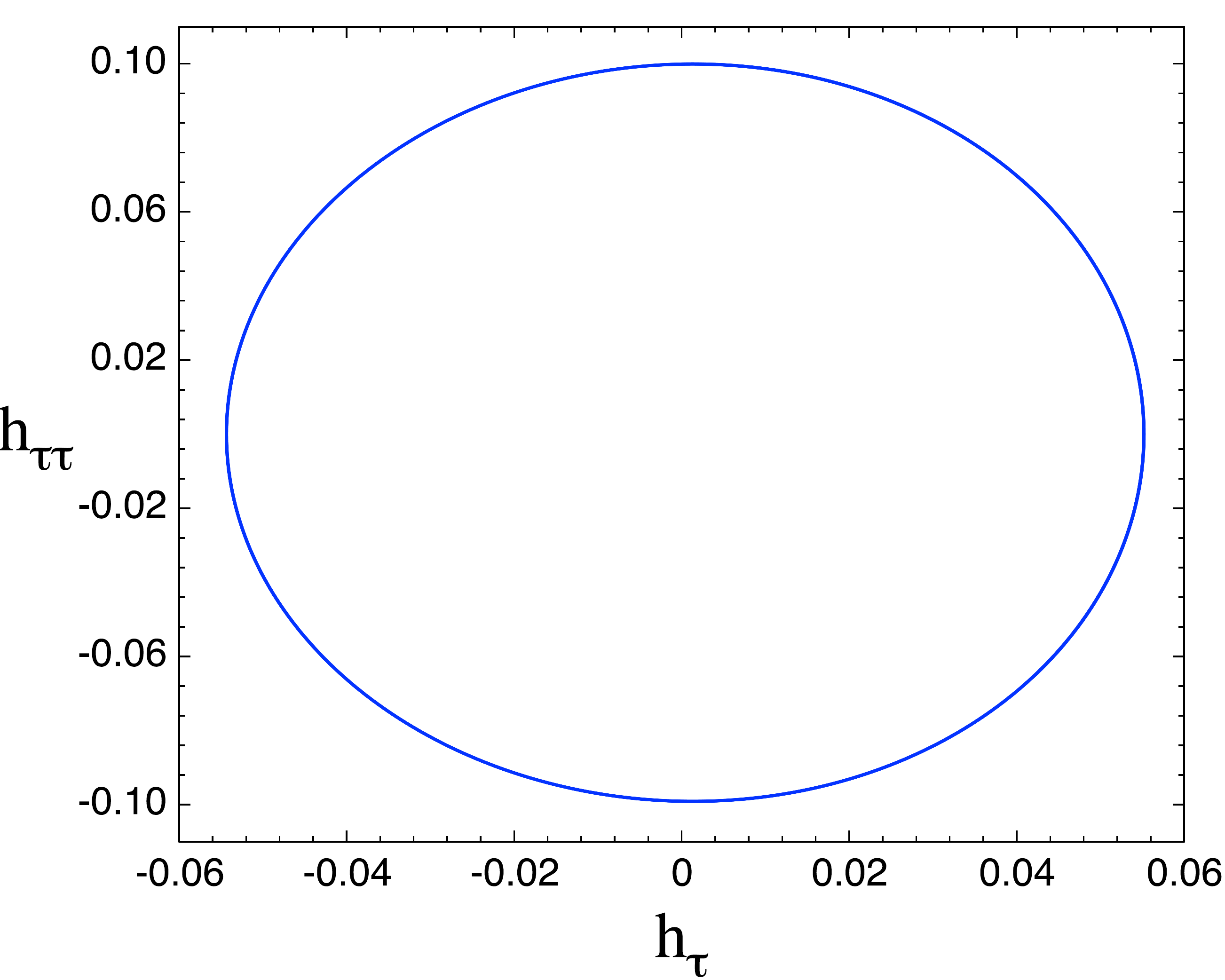}}
	\caption{Neutrally stable solution for $K=0.1$ and $\alpha=44$.}\label{fig:critical}
\end{figure}
\begin{figure}[ht]
	\centering
	\subfigure[Shock front density history]{
	\includegraphics[width=0.45\linewidth]{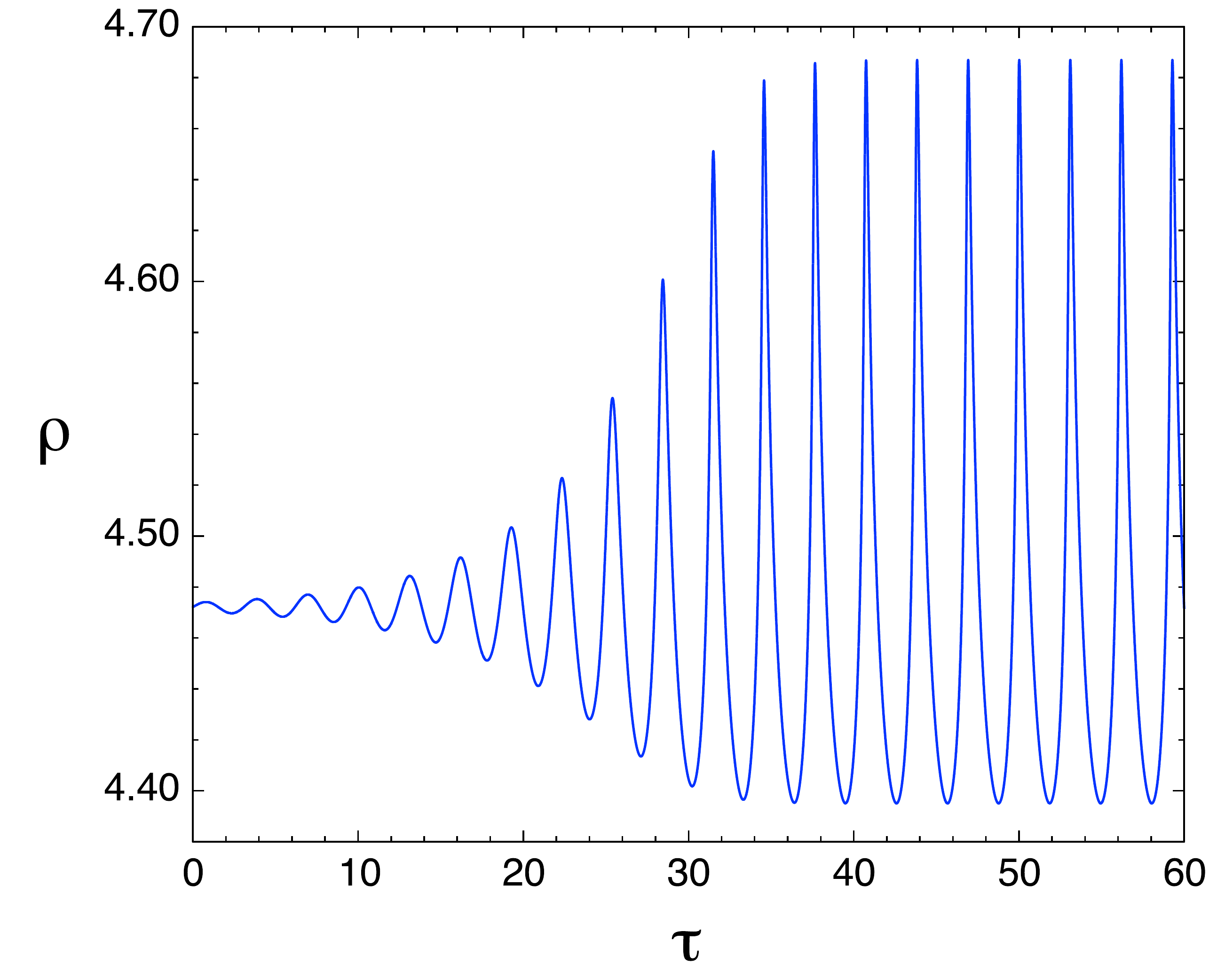}}
	\subfigure[Shock front acceleration vs. relative velocity.]{
	\includegraphics[width=0.45\linewidth]{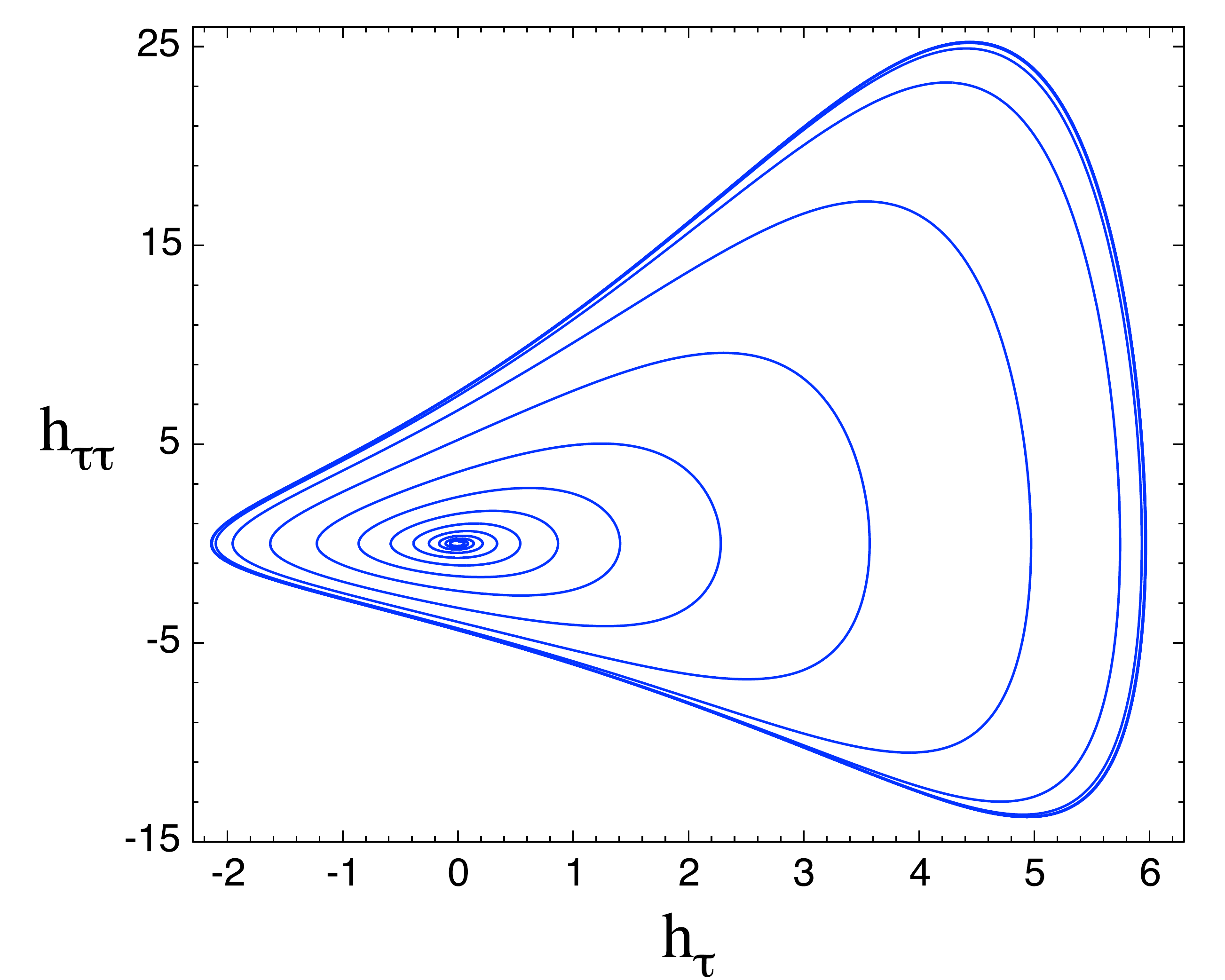}}
	\caption{Unstable solution for $K=0.1$ and $\alpha = 55$.}\label{fig:unstable}
\end{figure}
\begin{figure}[ht]
	\centering
	\subfigure[Shock front density history]{
	\includegraphics[width=0.45\linewidth]{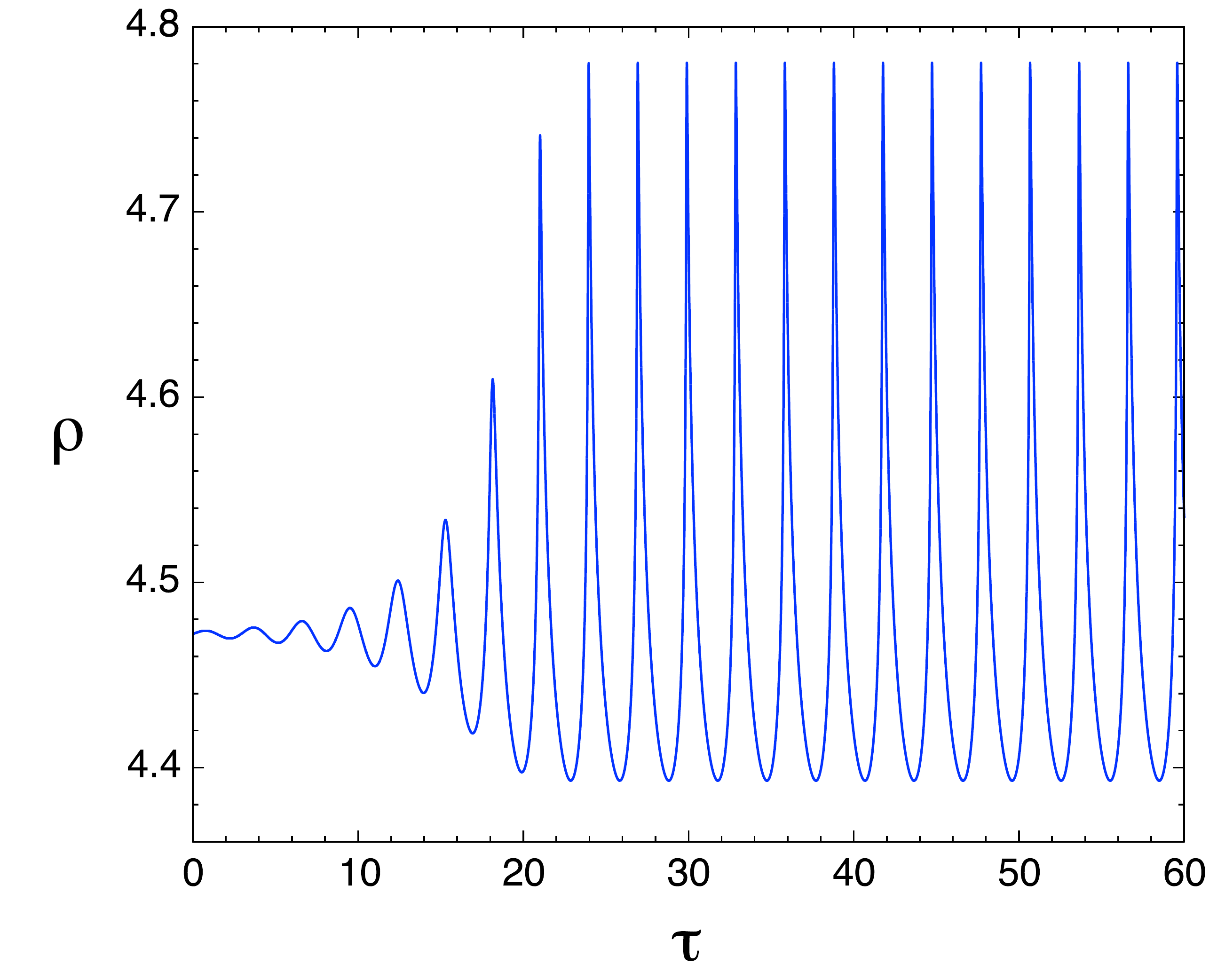}}
	\subfigure[Shock front acceleration vs. relative velocity.]{
	\includegraphics[width=0.45\linewidth]{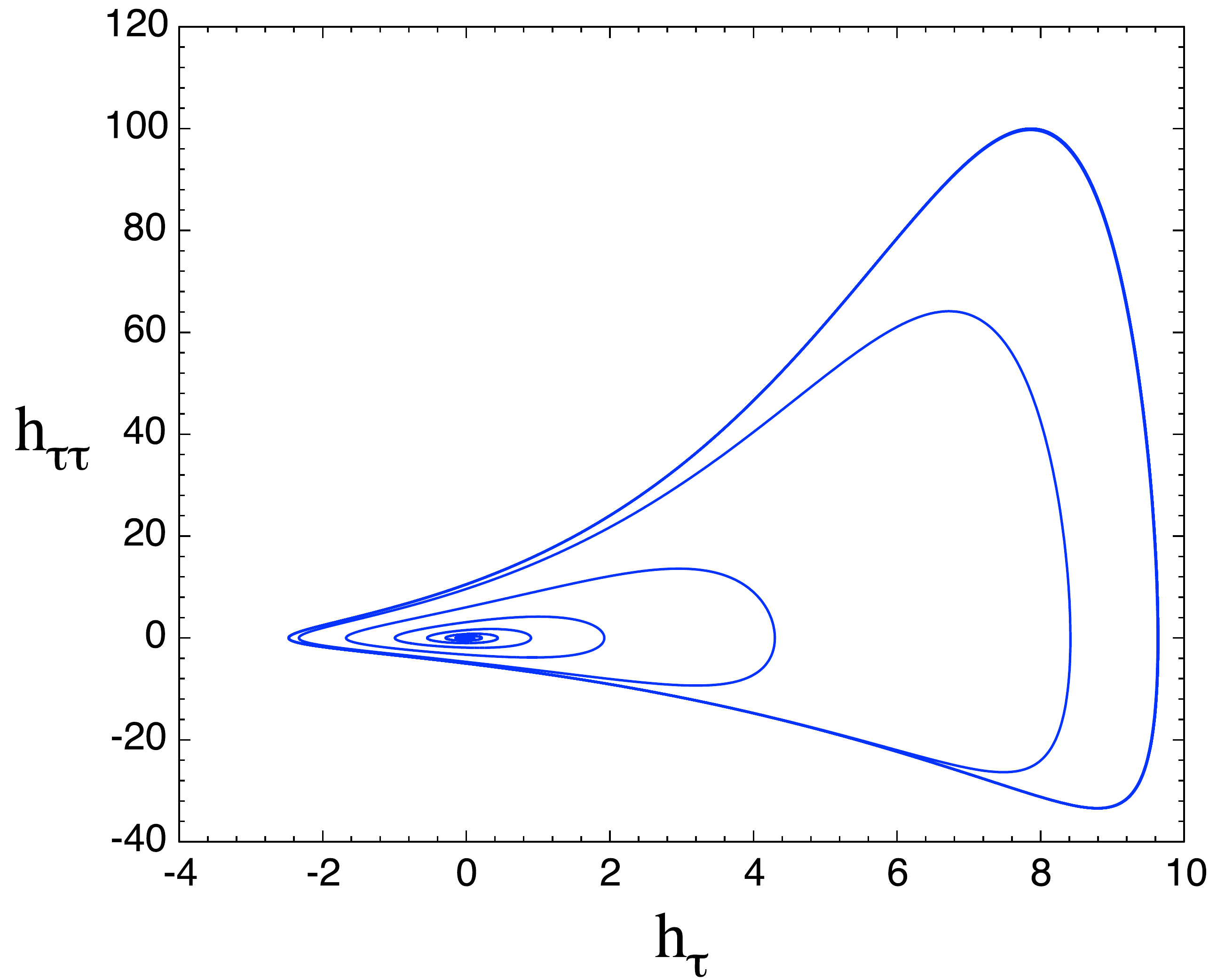}}
	\caption{(a) Increasingly unstable solution for $K=0.1$ and $\alpha = 63$, and (b) shock speed relative to steady wave phase portrait.}\label{fig:unstable.more}
\end{figure}

The parameters affecting stability were found to be the ratios of induction to reaction time $K$, the activation energy $\alpha$, and the reaction order $\nu$. The heat release parameter, $Q$ in \eqref{pressure.eq}, was not found to affect dynamics of instability. First, $K$ and $\nu$ were kept constant and $\alpha$ was varied. It was found that an increase in $\alpha$ promotes instability. For example, these solutions are plotted at a constant $K=0.1$ and $\nu=1/2$ with increasing $\alpha$ in Figs.~\ref{fig:stable} to~\ref{fig:unstable.more}. Figure~\ref{fig:stable} shows a stable solution, where the perturbations decays to the steady CJ detonation velocity. Further increasing $\alpha$ until the solution no longer decays shows a neutral stability point, where the perturbation does not decay nor amplify, see Fig.~\ref{fig:critical}. An additional increase in $\alpha$ gives rises to higher amplitude limit-cycles, see Fig.~\ref{fig:unstable}, where the amplitude initially grows exponentially then stabilizes in a periodic limit cycle. Increasing the activation energy even more, results in a higher amplitude and a higher frequency of oscillation, see Fig.~\ref{fig:unstable.more}. The evolution equation \eqref{evolution.eq} becomes unsolvable for increasing instability, due to the increasingly stiff behavior of the ODE.  

We then kept $\alpha$ and $K$ constant and found that decreasing the reaction order $\nu$, i.e., shortening the reaction length, results in unstable solutions. Similarly, keeping $\alpha$ and $\nu$ constant and increasing the induction to reaction time ratio, $K$, will also result in unstable solutions, provided the activation energy $\alpha$ is large enough.

\section{Linearized Oscillator and Neutral Stability Limit}
To predict stable and unstable solutions, the evolution equation~\eqref{evolution.eq} is linearized at the stability point $h_\tau=h_{\tau\tau}=0$, i.e., the idealized CJ detonation wave,
\begin{equation}
\label{linearized.evolution.eq}
\alpha_1 h_{\tau\tau\tau} +(\alpha_1+\alpha_2)h_{\tau\tau}+\alpha_3 h_\tau=0.
\end{equation}
The linearized equation is a damped ($\alpha_1+\alpha_2>0$) or amplified ($\alpha_1+\alpha_2<0$) harmonic oscillator. Since the initial shock front velocity perturbation is assumed to be small, the linearized system accurately predicts the early perturbation decay or growth in the nonlinear evolution equation~\eqref{evolution.eq}. Accordingly, the damping term's coefficient in the linearized equation~($\alpha_1$+$\alpha_2$ in \eqref{linearized.evolution.eq}) directly controls if energy is transfered into the reaction zone, as feedback, or out of the reaction zone, as attenuation. For stable solutions, a positive damping term, $\alpha_1+\alpha_2>0$ in the linearized evolution equation~\eqref{linearized.evolution.eq}, gives rise to the stability criteria
\begin{equation}
\label{stability.simplified}
\frac{1}{\epsilon}<\frac{2}{K(1-\nu)}+4. 
\end{equation}

\begin{figure}[ht]
\centering
\includegraphics[width=0.6\linewidth]{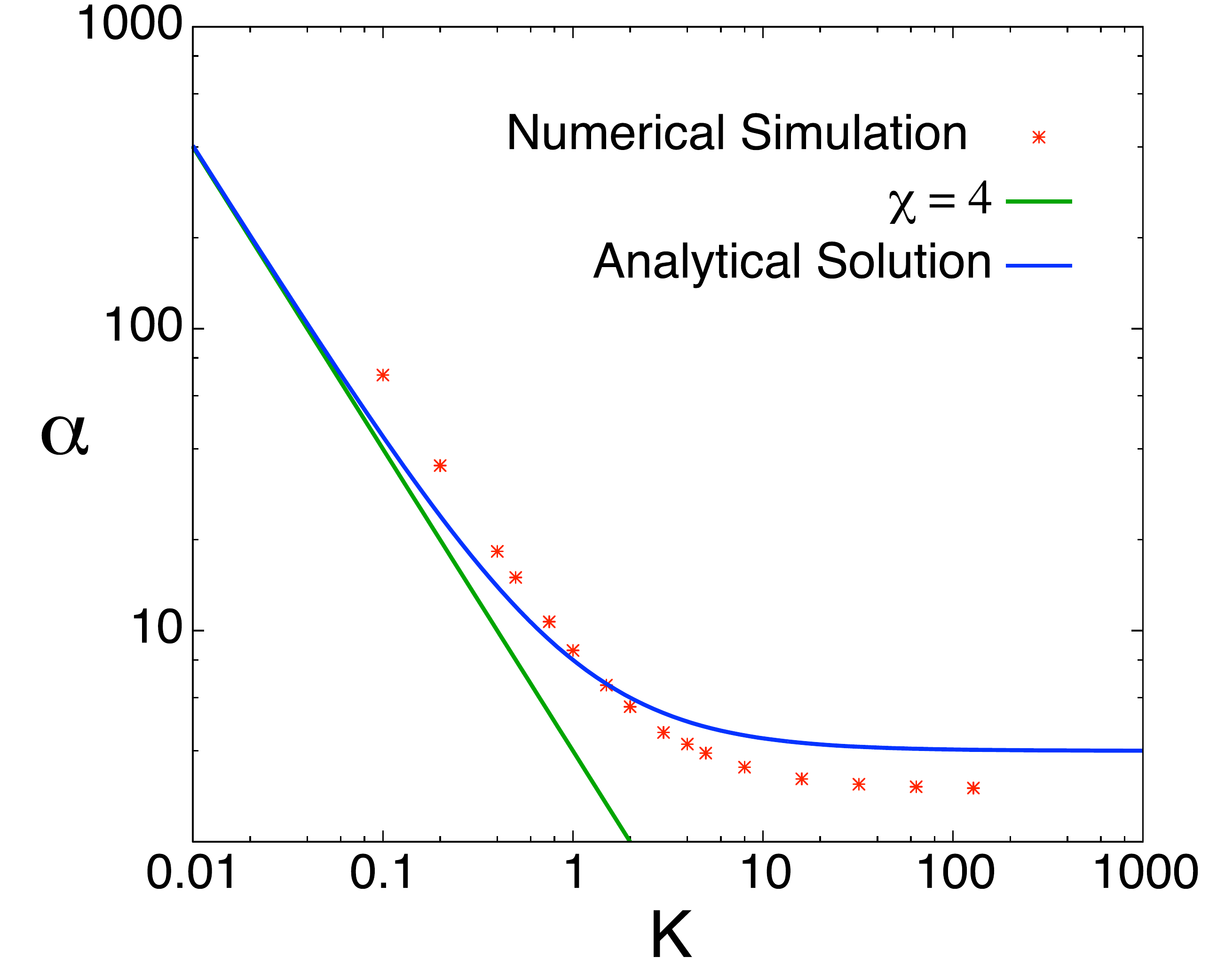}
\caption{Stability boundary for a reaction order of $\nu=1/2$ from the evolution equation in blue compared to numerical simulation in red dots\cite{tang_2013}.}
\label{fig:Stability_Boundary}
\end{figure}
This stability criteria agrees with numerical simulations, see Figure \ref{fig:Stability_Boundary}. At the limit of $K\ll1$, the product of activation energy times the ratio of induction time to reaction time is equal to a constant, the non-dimensional number
\begin{equation}
\chi=\frac{1}{\epsilon}K<4.
\end{equation}
In contrast, the limit of $K\gg1$ shows the stability to be uniquely controlled by the activation energy, ${\alpha<4}$. Somewhat surprisingly, we find that the analysis recovers the correct neutral stability boundary even when the conditions for its derivation ($K\ll1$) are no longer true, as can be seen in Fig.~\ref{fig:Stability_Boundary}.

\section{Stability Mechanism}
To identify the underlying mechanisms behind perturbation growth and decay, the damping term $(\alpha_1+\alpha_2)$ in~\eqref{linearized.evolution.eq} is traced back prior to calculations:
\begin{equation}
\label{stability.eq.root}
\underbrace{\int_{0}^{F(\tau)}\frac{\partial \rho}{\partial \tau}^{(1)}\diff n}_{\text{From induction zone}}
+\underbrace{\int_{F(\tau)}^{R(\tau)}\frac{\partial \rho}{\partial \tau}^{(1)}\diff n}_{\text{{From reaction zone}}}
-\frac{1}{\epsilon}\underbrace{\int_{\lambda_r^{(0)}=0}^{\lambda_r^{(0)}=1}\diff\bigg((\rho^{(0)}+\frac{\lambda_r^{(0)}}{4})F_\tau\bigg)}_{\text{From reaction zone}}< 0,
\end{equation}
where, recalling $F(\tau)$ is the induction length~\eqref{induction.length}, and $R(\tau)-F(\tau)$ is the reaction length. The second term of the left hand side can also be written as
\begin{equation}
\label{reaction.integral}
\int_{F(\tau)}^{R(\tau)}\frac{\partial \rho}{\partial \tau}^{(1)}\diff n=\frac{1}{K}\int_{0}^{1}\frac{1}{(1-\lambda_r^{(0)})^\nu}\frac{\partial \rho}{\partial \tau}^{(1)}\diff\lambda_r^{(0)},
\end{equation}
to illustrate the effects of the induction to reaction time ratio, $K$. 

The leading order element of \eqref{stability.eq.root} originates from the $O(\epsilon)$ evolution equation~\eqref{evolution.O1.eq}, and is responsible for perturbation amplification, where an increase in shock velocity, $h_\tau$, leads to a large decrease in induction length~\eqref{induction.length}. As a result, the flame speed, $F_\tau=h_{\tau\tau}e^{-h_\tau}$, begins accelerating forward, which increases the flux entering the reaction zone. Consequently, the pressure~\eqref{pressure.eq} in the reaction zone increases. Through the conservation equation~\eqref{conservation.eq}, the shock front must now accelerate forward to achieve equilibrium at the fully reacted point. The feedback results in a runaway acceleration of the shock front at leading order. Mathematically, this is expressed as $h_{\tau\tau}\propto h_\tau$ at the equilibrium point, which admits exploding solutions, since acceleration is proportional to velocity. The activation energy controls how sensitive the induction length is to changes in density through the rate equation~\eqref{rate.ind.normalized.eq}. For this reason, increasing the activation energy increases instabilities, as seen in Figs.~\ref{fig:stable} to~\ref{fig:unstable.more}.

Positive dampening/attenuation only emerges at higher order, namely $\partial/\partial t$ terms are no longer zero. Amplifications originating from the rear equilibrium point are now communicated along $n$ and $\tau$ coordinates, i.e., acoustic waves. The element leading to attenuation is due to $h_\tau$ in $\partial_\tau \rho^{(1)}$ in the first and second term of~\eqref{stability.eq.root}, i.e., shock acceleration. The resulting acoustic waves manifest themselves as gradients in the induction and reaction zone, sloped forwards compressing the media, and sloped backwards expanding the media. Mathematically, it is expressed as $h_{\tau\tau}\propto -h_\tau$ to get equilibrium at the fully reacted point. Where by itself admits stable exponential solutions decaying to $h_{\tau\tau}=h_\tau=0$.

The magnitude of the dampening term is a result of the gradients creating a net difference in density between the equilibrium point and the shock front, relative to the steady solution. Consequently, the length of both the induction and reaction zone will determine the net difference in density, the first and second term of \eqref{stability.eq.root}. From equation~\eqref{reaction.integral}, longer reactions, i.e., decreasing $K$, leads to stronger attenuation. If the total net difference in density between the equilibrium point and shock front, due to shock speed perturbation, is larger than the amplification mechanism as previously discussed, the third term of \eqref{stability.eq.root}, than the perturbations will decay, and vice versa. 

The unstable solutions reach periodic limit cycles, see Fig.~\ref{fig:unstable}~(b) and~\ref{fig:unstable.more}~(b), because the amplification mechanism balances out with the attenuation mechanism. This balance occurs due to the flame acceleration going to zero when the induction length goes to zero as shock speed increases. In other words, at the peak amplitude in Figs.~\ref{fig:unstable} and~\ref{fig:unstable.more}, the damping term ($\alpha_1+\alpha_2\exp({-h_\tau})$) in \eqref{evolution.eq} is effectively zero.

The oscillatory mode is from the same origin as the attenuation term but is instead due to the flame speed's rate of change in the reaction zone, coming from the second term of~\eqref{stability.eq.root}. The flame speed acceleration is of the form $F_{\tau\tau}=h_{\tau\tau\tau}\exp{-h_\tau}+...$, and acts as an equivalent inertial term that is analogous to the mass in a spring and damper system. Mathematically, this is can be expressed as $h_{\tau\tau\tau}\propto-h_\tau$ to achieve equilibrium, where on its own results in neutral oscillation. Longer reaction layers, i.e., lowering $K$, leads to a longer time for the perturbations from $F_{\tau\tau}$ to reach the shock front, resulting in a longer period of oscillation.

\section{Concluding Remarks}
A second order, nonlinear evolution equation is derived for pulsating Chapman-Jouguet detonation waves. The evolution equation predicts stable and unstable solutions. For longer reaction times compared to induction times, lowering the product of activation energy and ratio of induction to reaction time leads to stability. For shorter reaction times compared to induction times, lowering the activation energy alone leads to stability. All unstable numerical solutions calculated have a stable periodic limit cycle. However, increasing the activation energy and ratio of induction to reaction time well beyond the stability boundary leads to a progressively stiff evolution equation which becomes unsolvable with the current numerics used.

In the induction zone, a small perturbation leads to large change in induction time owing to a high activation energy, which in turn increases the flux entering the reaction zone, effectively amplifying the shock front further through the conservation law. The stability mechanism is higher order and originates from both the reaction and induction zone. A net density difference due to gradients from acoustics in the reaction and induction zone, controlled by their respective lengths, attenuates shock amplification. If the net increase is larger than the additional reaction zone flux then the detonation wave is stable, and vice versa. The dynamics observed with the evolution equation agree with numerical simulations as well as detonation dynamics in reactive Euler systems.
Although the present analysis was successful in explaining the instability mechanism of detonations and predicting the neutral stability boundary in terms of the $\chi$ parameter, it does not predict any subsequent period doubling bifurcation and chaotic dynamics previously observed for this system and the reactive Euler equations\cite{radulescu_2005,radulescu_2011,kasimov_2013}. Further work is required to address this problem.
\bibliography{master}
\bibliographystyle{Det_Symp}

\end{document}